\documentclass[letter]{article}

\usepackage[english]{babel}
\usepackage[utf8x]{inputenc}
\usepackage[T1]{fontenc}
\usepackage{setspace}

\usepackage[sort&compress,square,numbers]{natbib}

\usepackage[a4paper,top=3cm,bottom=2cm,left=3cm,right=3cm,marginparwidth=1.75cm]{geometry}

\usepackage{amsmath}
\usepackage{graphicx}
\usepackage{authblk}
\usepackage[colorinlistoftodos]{todonotes}
\usepackage[colorlinks=true, allcolors=blue]{hyperref}

\title{Elements and Principles for Characterizing Variation between Data Analyses}
\author[1]{Stephanie C. Hicks}
\author[1]{Roger D. Peng}
\affil[1]{Department of Biostatistics, Johns Hopkins Bloomberg School of Public Health}

\begin{document}
\maketitle


\begin{abstract}
The data revolution has led to an increased interest in the practice of data analysis. For a given problem, there can be significant or subtle differences in how a data analyst constructs or creates a data analysis, including differences in the choice of methods, tooling, and workflow. In addition, data analysts can prioritize (or not) certain objective characteristics in a data analysis, leading to differences in the quality or experience of the data analysis, such as an analysis that is more or less reproducible or an analysis that is more or less exhaustive. However, data analysts currently lack a formal mechanism to compare and contrast what makes analyses different from each other. To address this problem, we introduce a vocabulary to describe and characterize variation between data analyses. We denote this vocabulary as the elements and principles of data analysis, and we use them to describe the fundamental concepts for the practice and teaching of creating a data analysis. This leads to two insights: it suggests a formal mechanism to evaluate data analyses based on objective characteristics, and it provides a framework to teach students how to build data analyses. 
\end{abstract}

\section{Introduction}

The data revolution has led to an increased interest in the practice of \textit{data analysis} \cite{tuke:1962, tukeywilk1996, box1976, wild1994, chatfield1995, wildpfannkuch1999, cook2007, grol:wick:2014} and increased demand for training and education in this area \cite{Cleveland2001, nolanlang2010, asaundergrad2014, baumer2015, pwc-datascience, hardin2015, kaplan2018, hicks2018}. For a given problem, a data analyst makes analytic choices such as which methods, algorithms, computational tools, languages or workflow to use in a data analysis. However, even when using the same data to investigate the same question, previous work has shown that there can be significant variation in how data analysts build data analyses \cite{Silberzahn2018}, which has been shown to influence the results of the analysis. 

One approach to understanding how data analysts make these analytic choices is related to the field of cognitive science in which the data analysis process is characterized as a sensemaking task whereby theories or expectations are set and then compared to reality (data) \cite{grol:wick:2014}. Any discrepancies between the two are further examined and then theories are possibly modified. While this cognitive model is useful for describing the data analysis process, this process is typically not observed by outsiders. Having the ability to characterize differences in the observed outputs (agnostic to the analyst who built the analysis) has benefits in particular for teaching data analysis, because it allows students and teachers to discuss the impact of the analytic choices made and discuss how to improve the analysis. Therefore, an alternative approach would be to characterize the observed outputs of the data analysis, so that individual analyses can be described and compared to other analyses in a concrete and agnostic manner. However, we currently lack a formal mechanism to describe these differences in the analytic choices and observed outputs in an agnostic manner. 

In addition to the analytic choices, a data analyst can prioritize (or not) objective characteristics in a data analysis, leading to differences in the quality or experience of the data analysis. An example of such an objective characteristic is when a data analyst prioritizes exhaustively checking a set of assumptions of a method instead of making a more modest effort. The result of the data analysis might not change, but the experience from the audience is changed from being less confident to more confident in the results from this part of the analysis as the degree of exhaustively checking the assumptions increases. However, we currently lack a formal mechanism, or vocabulary of objective characteristics, to compare and contrast what makes analyses different from each other.

Other fields, such as art or music, have overcome similar challenges by defining a vocabulary that can be used to characterize the variation between different pieces of art. More formally, an artist can create a piece of art using the \textit{elements} and \textit{principles} specific to that area. The elements of art include color, line, shape, form, and texture \cite{nga-elements}; and the principles of art are the means by which the artist uses to compose or to organize the elements within a work of art \cite{marder2018}. For example, an artist can use the principle of contrast (or emphasis) to combine elements in a way that stresses the differences between those elements, such as combining two contrasting colors, black and white. The principles of art, by themselves, are not used to evaluate a piece of art, but they are meant to be objective characteristics that can describe the variation between pieces of art. 

Here, we introduce a vocabulary to describe and characterize variation between the observed outputs of data analyses. We denote this vocabulary as the \textit{elements and principles of data analysis}, and we use them to describe the fundamental concepts for the practice and teaching of data analysis. Briefly, the elements of an analysis are the individual basic components of the analysis that, when assembled together by the analyst, make up the entire analysis (Section~\ref{sec-elements}). The principles of the analysis are prioritized qualities or characteristics that are relevant to the analysis, as a whole or individual components, and that can be objectively observed or measured (Section~\ref{sec-principles}). Using short vignettes (Section~\ref{sec-vignettes}), we argue this vocabulary leads to two insights: it suggests a formal mechanism to evaluate data analyses based on these objective characteristics, and it provides a framework to teach students how to make analytic choices when building data analyses (Section~\ref{sec-discussion}).

\section{Elements of data analysis}
\label{sec-elements}
The \textit{elements} of a data analysis are the fundamental components of a data analysis used by the data analyst: code, code comments, data visualization, non-data visualization, narrative text, summary statistics, tables, and statistical models or computational algorithms \cite{Breiman2001} (Table~\ref{table-elements}). 

\begin{table}[ht!]
  \caption{\textbf{Elements of a data analysis.} This table describes eight elements that are used by the data analyst to build the data analysis.}
    \label{table-elements}
  \begin{tabular}{p{3cm}|p{11cm}}
  \hline
  \textbf{Element} & \textbf{Description} \\ 
  \hline 
  Narrative text  & Expository phrases or sentences that describe what is happening in the data analysis in a human readable format \\
  \hline
 Code &  A series of programmatic instructions to execute a particular programming or scripting language \\
  \hline
 Code comment & Non-executable code or text near or inline with code that describes the expected action/result of the surrounding code or provides context \\
  \hline
 Data visualization & A plot, figure or graph illustrating a visual representation of the data. \\
  \hline
 Narrative diagram & A diagram or flowchart without data \\ 
  \hline 
 Summary statistics & Numerical quantities derived from the data, such as the mean, standard deviation, etc. \\
 \hline 
 Table & An ordered arrangement of data or summaries of data in rows and columns \\
  \hline 
 Statistical model or computational algorithm & Mathematical model or algorithm concerning the underlying data phenomena or data-generation process, predictive ability, or computational algorithm \\
  \hline
  \end{tabular}

\end{table}

Code and code comments are two of the most commonly used elements by the data analyst to describe the executable programmatic instructions for a set of operations or computations and the non-executable instructions that describe the action or result of the surrounding code. These can be an entire line, multiple lines or a short snippet. Examples of code include defining variables or writing functions. Code comments and narrative text are related because they both can include expository phrases or sentences that describe what is happening in the data analysis in a human readable format. However, the difference between the two is the code comment has a symbol in front of the narrative text, which instructs the document container to not execute this element. In addition, there are two types of visualizations elements used in data analysis, data visualization and non-data visualization, where the former can be a plot, figure or graph illustrating a visual representation of the data and the latter can be a figure relevant to the data analysis but does not necessarily contain data, such as a diagram or flowchart. 

There are two types of summary elements of a data analysis: summary statistics and tables, where the former are one (or more than one) dimensional numerical quantities derived from the data, such as mean or standard deviation, while the latter is an ordered arrangement of either data or summaries of the data into a row and column format. The last element of a data analysis is the statistical model or computational algorithm, which an analyst can use to investigate the data-generation process or predictive ability of different mathematical models or algorithms. 

In addition to these elements, there are also \textit{contextual inputs} to the data analysis, such as the main question or problem being addressed, the data, the choice of programming language to use, the audience, and the document or container for the analysis, such as Jupyter or R Notebooks. We do not include these as elements of data analysis, because these inputs are not necessarily decided or fundamentally modified by the analyst. Often an upstream entity such as a manager at a company, a collaborator at a university or scientific institute, or an educator in the classroom provides the framework for these contextual inputs. However, we note that often the data analyst will be expected to decide or contribute to these contextual inputs. In addition, it may be the analyst’s job to provide feedback on some of these inputs in order to further refine or modify them. For example, an analyst may be aware that a specific programming language is more appropriate for a planned analysis than the currently selected one.

\begin{figure}[!t]
    \centering
    \includegraphics[width=0.9\textwidth]{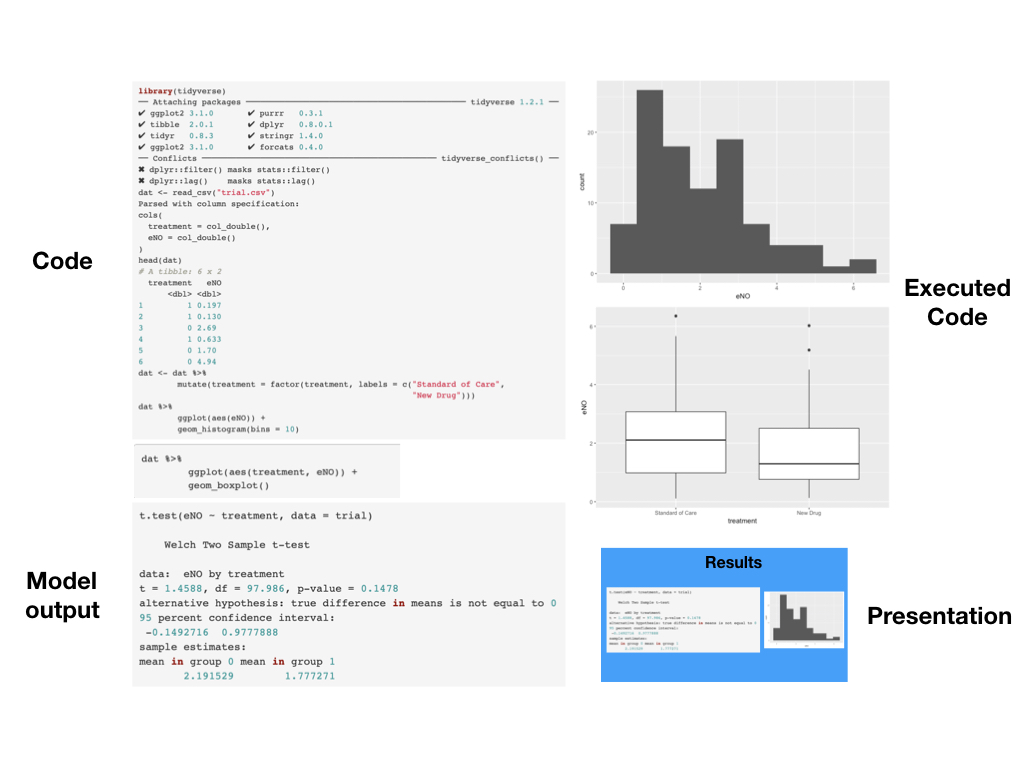}
    \caption{Sample analytic container, analytic product, and analytic presentation.}
    \label{fig:simpleexample}
\end{figure}

Finally, an analysis will usually result in potentially three basic outputs (Figure~\ref{fig:simpleexample}). The first is the analysis itself, which we imagine as living in an \textit{analytic container} which might be a set of files including a Jupyter notebook or R Markdown document, a dataset, and a set of ancillary code files. The analytic container is essentially the ``source code'' of the analysis and it is the basis for making modifications to the analysis and for reproducing its findings. In addition to the container, there will usually be an \textit{analytic product}, which is the executed version of the analysis in the analytic container, containing the executed code producing the results and output that the analyst chooses to include, which might be a PDF document or HTML file. Finally, the analyst will often produce an \textit{analytic presentation}, which might be a slide deck, PDF document, or other presentation format, which is the primary means by which the data analysis is communicated to the audience. Elements included in the analytic presentation may be derived from the analytic container, analytic product, or elsewhere.

\section{Principles of data analysis}
\label{sec-principles}
The \textit{principles} illustrated by a data analysis are prioritized qualities or characteristics that are relevant to the analysis, as a whole or individual components, and that can be objectively observed or measured. Their presence (or absence) in the analysis is not dependent on the characteristics of the audience viewing the analysis, but rather the relative weight assigned to each principle by the analyst can be highly dependent on the audience’s needs. In addition, the weighting of the principles by the analyst can be influenced by outside constraints or resources, such as time, budget, or access to individuals to ask context-specific questions, that can impose restrictions on the analysis. The weighting of the principles \textit{per se} is not meant to convey a value judgment with respect to the overall quality of the data analysis. Rather, the requirement is that multiple people viewing an analysis could reasonably agree on the fact that an analysis gives high or low weight to certain principles. In Section~\ref{sec-vignettes} we describe some hypothetical data analyses that demonstrate how the various principles can be weighted differently. Next, we describe six principles that we believe are informative for characterizing variation between data analyses.

\begin{figure}[ht!]
  \centering\includegraphics[width=.75\textwidth]{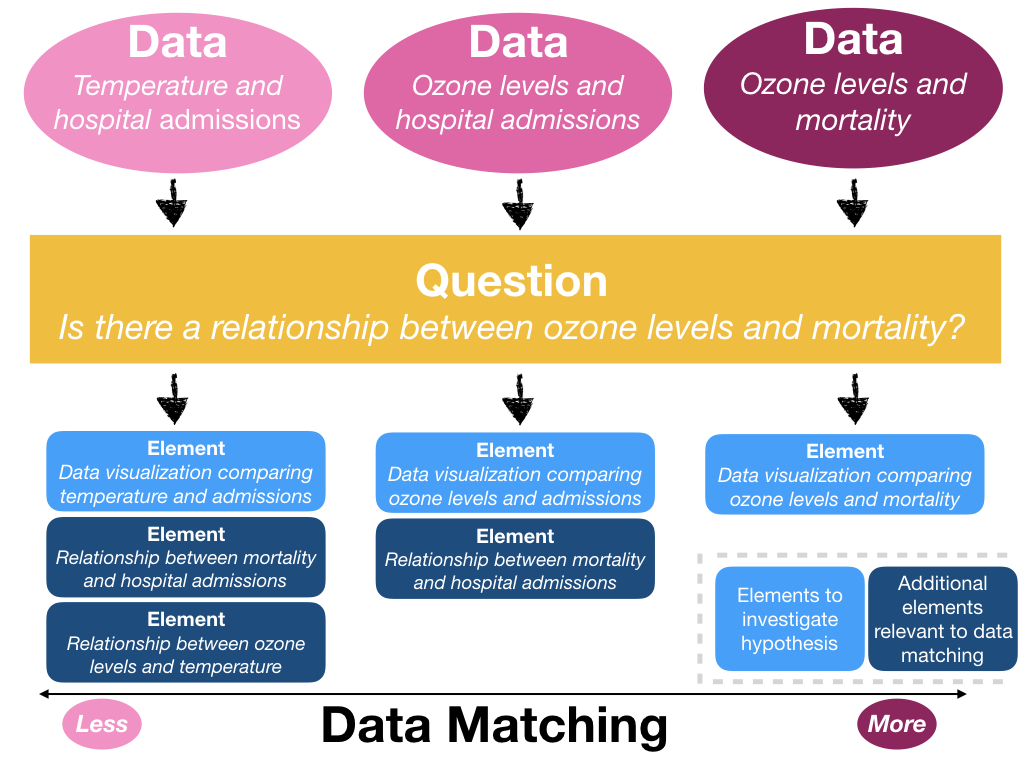}
  \caption{\textbf{The data matching principle of data analysis.} Data analyses with high data matching have data readily measured or available to the analyst that directly matches the data needed to investigate a question or problem with data analytic elements. In contrast, a question may concern quantities that cannot be directly measured or are not available to the analyst. In this case, data matched to the question may be surrogates or covariates to the underlying data phenomena that may need additional elements to describe how well the surrogate data is related to the underlying data phenomena to investigate the main question. }
  \label{fig-data-matching}
\end{figure}

\vspace{0.5em}\noindent\textbf{Data Matching}. Data analyses with high \textit{data matching} have data readily measured or available to the analyst that directly matches the data needed to investigate a question with data analytic elements (Figure~\ref{fig-data-matching}). In contrast, a question may concern quantities that cannot be directly measured or are not available to the analyst. In this case, data matched to the question may be surrogates or covariates to the underlying data phenomena. While we consider the main question and the data to be contextual inputs to the data analysis, we consider this a principle of data analysis because the analyst selects data analytic elements that are used to investigate the question, which depends on how well the data are matched. If the data are poorly matched, the analyst will not only need to investigate the main question with one set of data analytic elements, but also will need to use additional elements that describe how well the surrogate data is related to the underlying data phenomena to investigate the main question. 

It is important to note that questions can be more or less specific, which will impose strong or weak constraints on the range of data matching to the question. Highly specific questions tend to induce strong constraints to investigate with data analytic elements. Less specific questions emit a large range of potential data to investigate the question. Data that can be readily measured or are available to the analyst to directly address a specific question results in high data matching, but depending on the problem specificity, can result in a narrow or broad set of data to consider. 

\begin{figure}[t!]
  \centering\includegraphics[width=.75\textwidth]{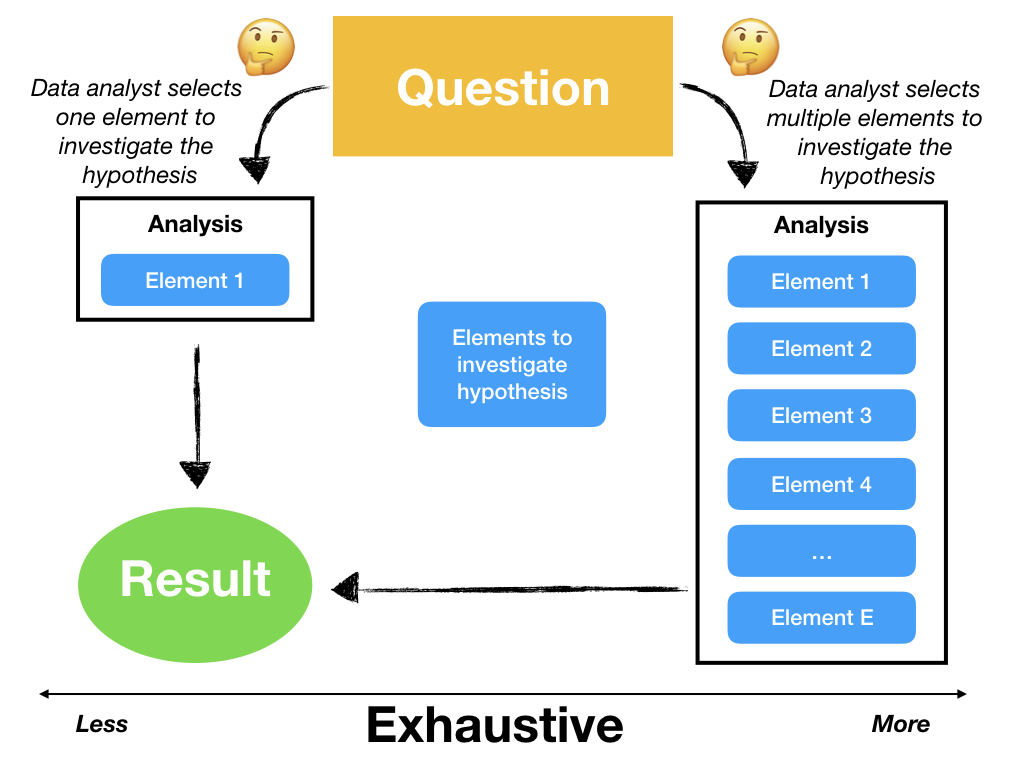}
  \caption{\textbf{The exhaustive principle of data analysis.} An analysis is exhaustive if specific questions are addressed using multiple, complementary elements. For a given question, the analyst can select an element or set of complementary elements to investigate the question. The more complementary elements that are used, the more exhaustive the analysis is, which provides a more complete picture of the evidence in the data than any single element. }
  \label{fig-exhaustive}
\end{figure}

\vspace{0.5em}\noindent\textbf{Exhaustive}. An analysis is \textit{exhaustive} if specific questions are addressed using multiple, complementary elements (Figure~\ref{fig-exhaustive}). For example, using a $2\times 2$ table, a scatter plot, and a correlation coefficient are three different elements that could be used to address the question if two predictors are correlated. Analysts that are exhaustive in their approach use complementary tools or methods to address the same question, knowing that each given tool reveals some aspects of the data but obscures other aspects. As a result, the combination of elements used may provide a more complete picture of the evidence in the data than any single element.

\begin{figure}[ht!]
  \centering\includegraphics[width=.75\textwidth]{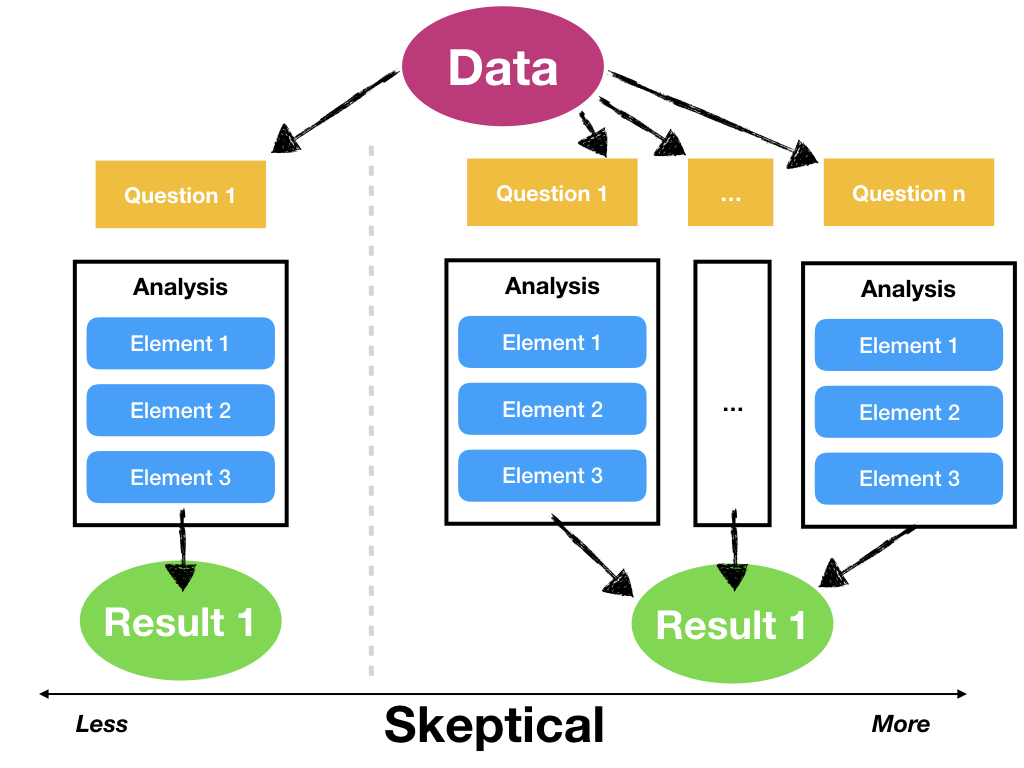}
  \caption{\textbf{The skeptical principle of data analysis.} An analysis is skeptical if multiple, related questions or alternative explanations of observed phenomena are considered using the same data and offer consistency of the data with these alternative explanations. In contrast, analyses that do not consider alternate explanations have no skepticism.}
  \label{fig-skeptical}
\end{figure}

\vspace{0.5em}\noindent\textbf{Skeptical}. An analysis is \textit{skeptical} if multiple, related questions are considered using the same data (Figure~\ref{fig-skeptical}). Analyses, to varying extents, consider alternative explanations of observed phenomena and evaluate the consistency of the data with these alternative explanations. Analyses that do not consider alternate explanations have no skepticism. For example, to examine the relationship between a predictor X and an outcome Y, an analyst may choose to use different models containing different sets of predictors that might potentially confound that relationship. Each of these different models represents a different but related question about the X-Y relationship. A separate question that arises is whether the configuration of alternative explanations are relevant to the problem at hand. However, often that question can only be resolved using contextual information that is outside the data. 

The need for more or less skepticism in a data analysis is typically governed by outside circumstances and the context in which the analysis sits. Analysis that may have large impacts or result in significant monetary costs will typically be subject to detailed scrutiny. In July 2000, the Health Effects Institute published a reanalysis of the Harvard Six Cities Study, which was a seminal air pollution study that showed significant associations between air pollution and mortality. Because of the potential regulatory impact of the study, HEI commissioned an independent set of investigators to reproduce the findings and conduct a series of sensitivity analyses~\citep{krew:burn:gold:hoov:2000}. The result was a nearly 300 page volume where the data and findings were subject to intense skepticism and every alternative hypothesis was examined. 

There are other instances when skepticism in the form of alternate explanations is not warranted in the analysis. For example, with an explicitly planned and rigorously-conducted clinical trial, the reported analysis will typically reflect only what was pre-specified in the trial protocol. Other analyses may be presented in a paper but they will be explicitly labeled as secondary. For example, in a large clinical trial studying the effect of a pest management intervention on asthma outcomes~\cite{matsui2017effect}, the reported analysis is ultimately a simple comparison of asthma symptoms in two groups. Some other secondary analyses are presented but they do not directly address the primary comparison. Such an analysis is acceptable here because of the strict pre-specification of the analysis and because of the standards and practices that the community has developed regarding the reporting of clinical trials.

\begin{figure}[t!]
  \centering\includegraphics[width=.75\textwidth]{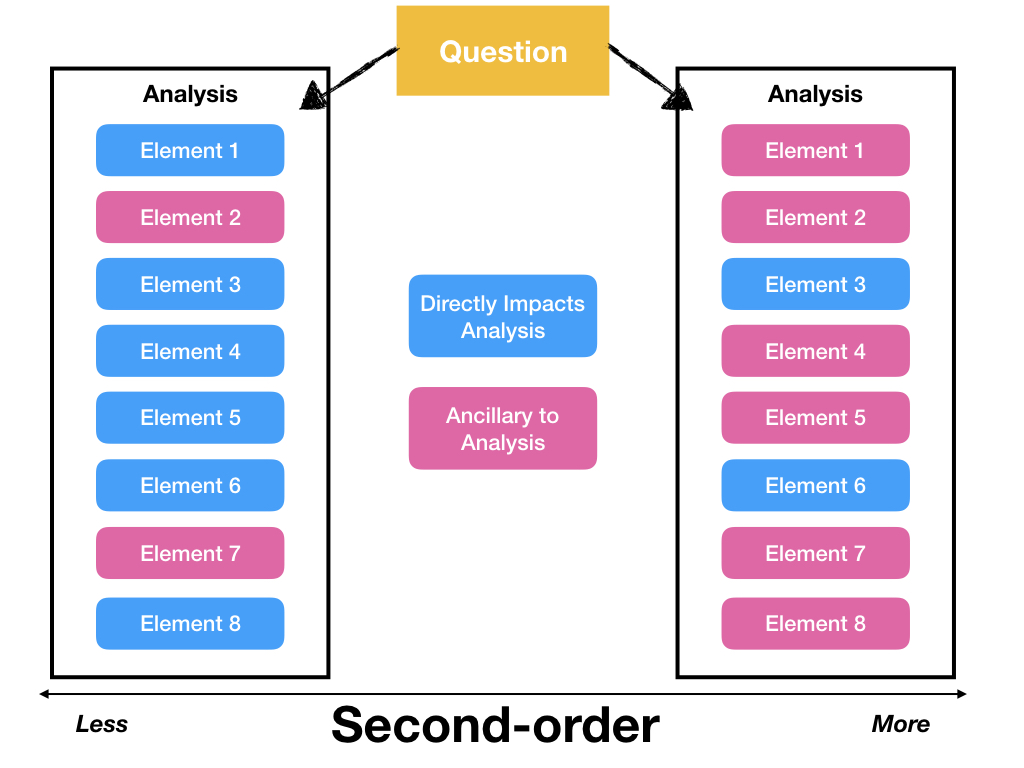}
  \caption{\textbf{The second-order principle of data analysis.} An analysis is second-order if it includes ancillary elements that do not directly address the primary question but give important context to the analysis. Examples of ancillary elements could be background information of how the data were collected, and expository explanations or analyses comparing different statistical methods or software packages. While these details may be of interest and provide useful background, they likely do not directly influence the analysis itself.
}
  \label{fig-second-order}
\end{figure}

\vspace{0.5em}\noindent\textbf{Second-Order}. An analysis is \textit{second-order} if it includes elements that do not directly address the primary question, but give important context or supporting information to the analysis (Figure~\ref{fig-second-order}). Any given analysis will contain elements that directly contribute to the results or conclusions, as well as some elements that provide background or context or are needed for other reasons, such as if the data are less well matched to the investigation of the question (Figure~\ref{fig-data-matching}). Second-order analyses contain more of these background/contextual elements in the analysis, for better or for worse. For example, in presenting an analysis of data collected from a new type of machine, one may include details of who manufactured the machine, or why it was built, or how it operates. Often, in studies where data are collected in the field, such as in people's homes, field workers can relay important details about the circumstances under which the data were collected. In both examples, these details may be of interest and provide useful background, but they may not directly influence the analysis itself. Rather, they may play a role in interpreting the results and evaluating the strength of the evidence.

\begin{figure}[ht!]
  \centering\includegraphics[width=.75\textwidth]{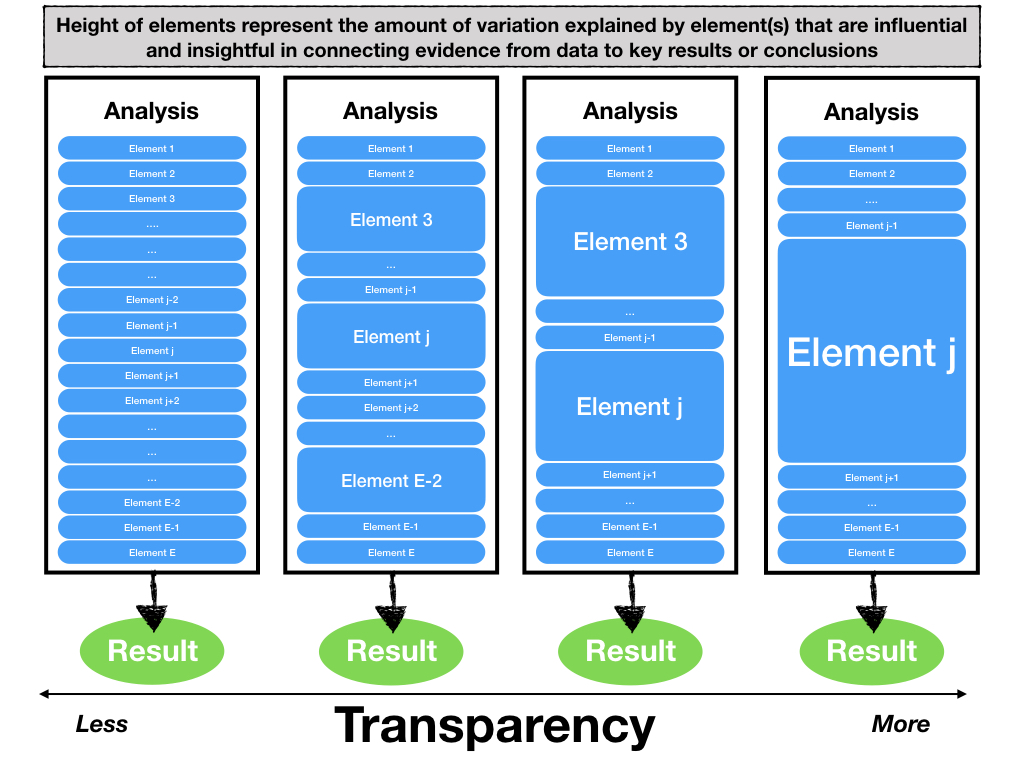}
  \caption{\textbf{The transparency principle of data analysis.} Transparent analyses present an element or set of elements summarizing or visualizing data that are influential in explaining how the underlying data phenomena or data-generation process connects to any key output, results, or conclusions. While the totality of an analysis may be complex and involve a long sequence of steps, transparent analyses extract one or a few elements from the analysis that summarize or visualize key pieces of evidence in the data that explain the most “variation” or are most influential to understanding the key results or conclusion. }
  \label{fig-transparent}
\end{figure}

\vspace{0.5em}\noindent\textbf{Transparent}. \textit{Transparent} analyses present an element or subset of elements summarizing or visualizing data that are influential in explaining how the underlying data phenomena or data-generation process connects to any key output, results, or conclusions (Figure~\ref{fig-transparent}). While the totality of an analysis may be complex and involve a long sequence of steps, transparent analyses extract one or a few elements from the analysis that summarize or visualize key pieces of evidence in the data that explain the most “variation” or are most influential to understanding the key results or conclusion. One aspect of being transparent is showing the approximate mechanism by which the data inform the results or conclusion.  

\begin{figure}[ht!]
  \centering\includegraphics[width=.75\textwidth]{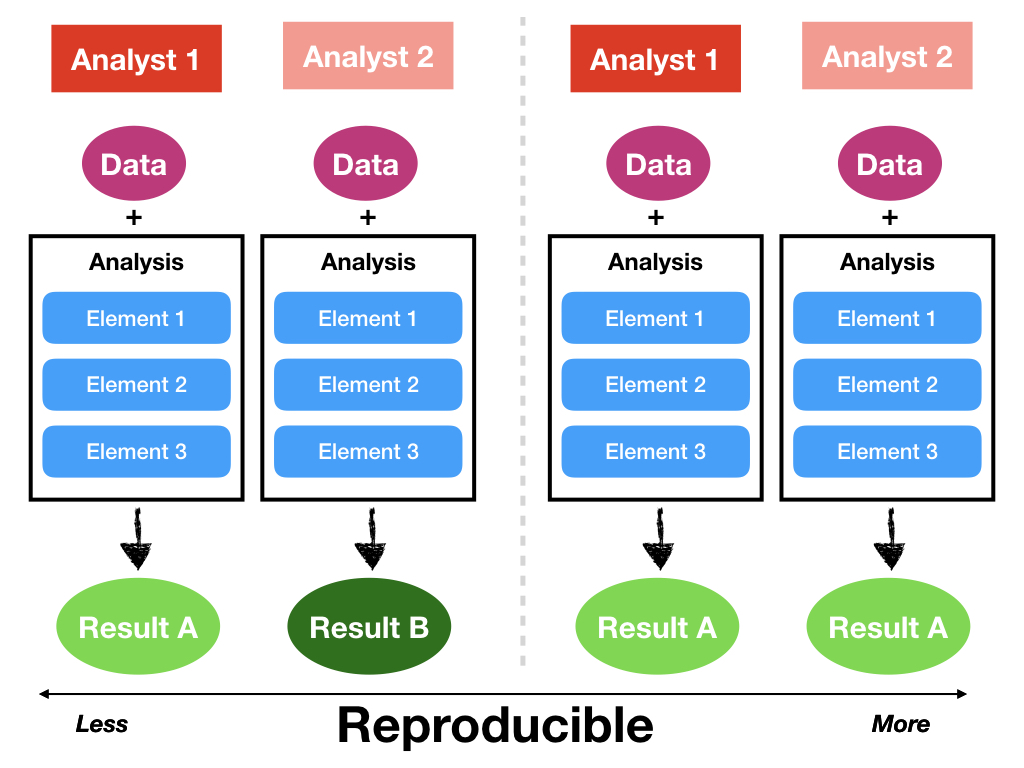}
  \caption{\textbf{The reproducible principle of data analysis.} An analysis is reproducible if someone who is not the original analyst (Analyst 2) can take the same data and the same elements of the data analysis and produce the exact same results as the original analyst (Analyst 1). In contrast, analyses that conclude in different results are less reproducible.}
  \label{fig-reproducible}
\end{figure}

\vspace{0.5em}\noindent\textbf{Reproducible}. An analysis is \textit{reproducible} if someone who is not the original analyst can take the published code and data and compute the same results as the original analyst (Figure~\ref{fig-reproducible}). In the terminology of our framework, given the elements of the data analysis, we can produce the exact same results of the analysis. Critical to reproducibility is the availability of the analytic container to others who may wish to re-examine the results. For example, analyses that integrate literate programming \cite{Knuth1984} in the analytic container make data analysis more reproducible \cite{Vassilev2016-literate}. Another consideration is that it may not be possible for businesses, such as those in the finance industry, to make available entire analytic containers for proprietary or financial reasons. In contrast, analytic containers that are integrated as part of the analytic product or analytic presentation are by definition more reproducible. Finally, much has been written about reproducibility and its inherent importance in science, so we do not repeat that here~\cite{peng:2011}. We simply add that reproducibility (or lack thereof) is usually easily verified and is not dependent on the characteristics of the audience viewing the analysis. Reproducibility also speaks to the coherence of the workflow in the analysis in that the workflow should show how the data are transformed to eventually become results.

\section{Vignettes}
\label{sec-vignettes}

To make these ideas more concrete, we provide four vignettes in this section where we describe how a data analysis could invoke or not invoke certain principles of data analysis.  

\subsection*{Vignette 1}

\textit{Background}. Roger is interested in understanding the relationship between outdoor air pollution concentrations and population health. However, monitoring of air pollution is expensive and time-consuming, and so he first develops a prediction model for predicting air pollution levels where there is no existing monitoring data. He then relates these predicted air pollution concentrations to respiratory disease hospitalization rates provided by a local insurance company that insures the majority of residents.

\textit{Analysis}. Using available data on air pollution concentrations as well as 20 other variables that he thinks would be predictive of pollution levels (e.g. temperature, wind speed, distance to road, traffic counts, etc.), he fits a linear model using measured monitor-level pollution concentrations as the outcome. Once the analysis is complete, he includes all of the code and the writeup in a Jupyter Notebook. The document and all the corresponding data are uploaded to GitHub and are made publicly available. He then runs a generalized linear model (GLM) with numbers of hospitalizations for respiratory diseases as the outcome and predicted air pollution concentration as the key predictor. Potential confounders are adjusted for directly in the regression model. From this model he obtains an estimate of the relative risk of hospitalization associated with predicted air pollution concentrations. To test the sensitivity of his findings to his initial model, he fits a series of additional models using different functional forms on the various confounders. He puts the health risk modeling code on GitHub but due to privacy concerns is unable to make the hospitalization data available.

\textit{Results}. In his analysis report, he indicates that in the pollution prediction model, temperature had a large coefficient in the model and was statistically significant. He further reports that a 1 degree increase in temperature was associated with a 2.5 unit increase in pollution. Other coefficients that were statistically significant were the coefficients for distance to road and wind speed. He also reports the percent increase in respiratory hospitalizations per 10 unit increase in pollution along with estimates obtained from some of the alternate models.

\subsubsection*{Mapping to Principles}

The stated goal of this analysis is to build a prediction model for predicting unobserved levels of air pollution and for assessing pollution's relationship to population health. A generalized linear model is fit and then the coefficients of the model are interpreted. 

 \begin{itemize}
    \item \textbf{Matching to the Data}: The data appear highly appropriate for addressing the problem of building a prediction model for pollution. Observed monitoring data are available for the outcome and 20 covariates that potentially related to pollution are used as predictors.
    \item \textbf{Exhaustive}: There is little evidence of exhaustiveness in this report. There is no attempt to implement alternative elements to see if additional insights can be gained. Essentially, one model was fit and the results reported.
    \item \textbf{Skeptical}: There is some skepticism in the analysis as multiple alternative models were explored, resulting in series of parameter estimates.
    \item \textbf{Second-order}: No second-order details are provided in the summary, such as background information about air pollution exposure or its relationship to health.
    \item \textbf{Transparent}: The level of transparency is low as there is no data visualization or data summary included in the analysis that highlights the evidence in the data or data-generation process that reveal how the reported results are influenced by features in the data.
   \item \textbf{Reproducible}: The code and data are made available on GitHub and the code for implementing the model is organized in a Jupyter Notebook. That aspect of the analysis therefore is reproducible. However, the health data are not made available and so the health risk modeling cannot be reproduced.
\end{itemize}

Note that in the first part of the analysis the goal was to build a predictive model. However, the analyst ultimately reported the results as an inferential analysis. The principles of data analysis do not characterize the validity or success of the analysis, nor the strength or quality of evidence for the hypothesis of interest. However, we propose a framework for how the elements and principles of data analysis might be used for these ideas in Section~\ref{sec-discussion}.

\subsection*{Vignette 2}

\textit{Background}. Stephanie works as a data scientist at a small startup company that sells widgets over the Internet through an online store on the company’s web site. One day, the CEO comes by Stephanie’s desk and asks her how many customers have typically shown up at the store’s website each day for the past month. The CEO waits by Stephanie’s desk for the answer.

\textit{Analysis}. Stephanie launches her statistical analysis software and, typing directly into the software’s console, immediately pulls records from the company’s database for the past month. She then groups the records by day and tabulates the number of customers. From this daily tabulation she then calculates the mean and the median count. She then quickly produces a time series plot of the daily count of visitors to the web site over the past month.

\textit{Results}. Stephanie verbally reports the daily mean and median count to the CEO standing over her shoulder. While showing the results she briefly describes how the company's database system collects information about visitors and its various strengths and weaknesses. She also notes that in the past month the web site experienced some unexpected down time when it was inaccessible to the world for a few hours.

\subsubsection*{Mapping to Principles}

This scenario is a typical ``quick analysis'' that is often done under severe time constraints and where perhaps only an approximate answer is required. In such circumstances, there is often a limited ability to weight certain principles very heavily.
 \begin{itemize}
    \item \textbf{Matching to Data}: The data are essentially perfectly matched to the problem. The database tracks all visitors to the web site and the analysis used data directly from the database.
    \item \textbf{Exhaustive}: There is some exhaustiveness here as the analysis presented both the mean and the median (two different elements) as a summary of the typical number of customers per day.
    \item \textbf{Skeptical}: The analysis did not address any other hypotheses or questions.
    \item \textbf{Second-order}: Details about how the company's database operates and noting that the web site experienced some downtime are second order details. The information may impact the interpretation of the data, but does not imply that the summary statistic is incorrect and does not directly impact the analysis.
    \item \textbf{Transparent}: The analysis is fairly simple and as such is transparent. The addition of the time series plot increases the transparency of the analysis. 
    \item \textbf{Reproducible}: Given that the results were verbally relayed and that the analysis was conducted on the fly in the statistical software’s console, the analysis is not reproducible.
\end{itemize}

\section{Discussion}
\label{sec-discussion}
In developing the \textit{elements} and \textit{principles} of data analysis, our goal is to define a vocabulary to describe and to characterize variation between the observed outputs of data analyses in a manner that is not specifically tied to the science or the application underlying the analysis. While the elements are the building blocks for a data analysis, the principles can be wielded by the analyst to create data analyses with diversity in content and design. Being able to describe differences in this manner allows data analysts, who may be working in different fields or on disparate applications, to have a set of concepts that they can use to have meaningful discussions. The elements and principles therefore broaden the landscape of data analysts and allow people from different areas to converse about their work and have a distinct shared identity. It is important to reiterate that the inclusion or exclusion of certain elements or the weighting of different principles in a data analysis do not determine the overall quality or success of a data analysis. 

However, the elements and principles we have laid out here do not make for a complete framework for thinking about data analysis and leave a number of issues unresolved. One question might be if this framework can ``detect'' a dishonest or fraudulent analysis just based on the existence (or not) of certain principles in an analysis. If an analyst has produced an analysis with misleading evidence, this would be troubling, but such an outcome does not necessarily have a one-to-one relationship with dishonest intention. On the contrary, misleading evidence can arise from even the most honest of intentions. In particular, when sample sizes are small, models do not capture hidden relationships, there is significant measurement error, or for any number of other analytical reasons, evidence can lead us to believe something for which the opposite is true. However, such situations are not generally a result of fraud or intentional deceit. They are often a result of the natural iteration and incremental advancement of science. 

It is clear from the historical record that some analyses are not done with the best of intentions. The possibilities range from benign neglect, to misunderstandings about methodology, to outright fraud or intentional deceit. These analyses, unsavory as their origins may be, are nevertheless data analyses. Therefore, they should be describable according to the principles outlined here. The problem is there is no guarantee that dishonest analyses will always exhibit certain principles with specific weights. A truly wily analyst will be able to make an analysis exhibit certain principles, while still being misleading. 

Another unresolved issue with characterizing data analyses in our framework is the possible conflation of two important, but distinct, entities: the \textit{analysis} and the \textit{analyst} who conducted the analysis. For example, it is perhaps reasonable to think that an analysis presented by an inexperienced analyst might require more scrutiny than an analysis presented by a seasoned veteran. While both the analysis presented and the analyst behind it are important in the evaluation of the conclusions of an analysis, it is important to consider them separately. A key reason is because an analysis is presented to an audience, and therefore the audience by definition has all of the information about that analysis before them. Specific information about the data analyst is seldom available unless the audience has a personal relationship with the analyst or if the audience is very familiar with their work and has seen past examples. Therefore, requiring any characterization of an analysis to include information about the analyst would be entirely unworkable.

While this framework leaves some issues unresolved, it also points us in a few directions moving forward. The practice of data analysis is a rich, complicated, challenging topic because it involves not only the data analysis, but also requires the ability to characterize the success or the completeness of the analysis, and the strength or quality of evidence for the question of interest. For example, describing the variation between data analyses as variation in the weighting of different principles suggests a formal mechanism for evaluating the success of a data analysis. In particular, every data analysis has an audience that views the analysis and the \textit{audience} may have a different idea of how these various principles should be weighted. One audience may value reproducibility and exhaustiveness while another audience may value interactivity and brevity. Neither set of weightings is inherently correct or incorrect, but the success of an analysis may depend on how well-matched the analyst's weightings are to the audience's. Similarly, data analysts may be put in highly constrained situations where certain principles must be down-weighted or up-weighted. Regardless of the situation, an analyst who goes against the principle weightings that are demanded by the constraints may have some explaining to do. That said, audiences may be open to such explanation if the analyst can make a convincing argument in favor of a different set of weightings.

Another important area of consideration is the teaching of data analysis. The elements and principles may provide an efficient framework to teach students at scale how to analyze data, which is a significant problem given the demand for data analysis skills in the workforce. Because much data analysis education involves experiential learning with a mentor in a kind of apprenticeship model, there is a limit on how quickly students can learn the relevant skills while they gain experience. Having a formal language for describing different aspects of data analysis that does not require mimicking the actions of a teacher or time-consuming mentorship may serve to compress the education of data analysts and to increase the bandwidth for training. Furthermore, students and teachers can discuss different aspects of an analysis and debate which principles should be weighed more or less heavily.

Finally, the development of elements and principles for data analysis provides a foundation for a more general theory of data science. For example, one could imagine defining mathematical or set operators on the elements of data analysis and consider the ideas of independence and exchangeability. One could define the formal projection mapping between a given data analysis and a principle of data analysis. Alternatively, one could combine one or more elements into coherent activities to define \textit{units} or sections of a data analysis, such as the ``introduction'', ``setup'', ``data import'', ``data cleaning'', ``exploratory data analysis'', ``modeling'', ``evaulation'', ``communication'', and ``export'' units. There might not be a formal ordering of the units and the units can appear in a data analysis once, more than once or not at all. Then, a set of units can be assembled together into \textit{canonical forms} of data analyses, which are likely to vary across disciplines.

\section{Summary}

The demand for data analysis skills has grown significantly, leading to a re-examination of the practice and teaching of data analysis. Having a formal set of elements and principles for characterizing data analyses allows data analysts to describe their work in a manner that is not confounded by the specific application or area of study. Having concrete elements and principles also opens many doors for further exploration and formalization of the data analysis process. The benefits of developing elements and principles include setting the basis for a distinct identity for the field of data science and providing a potential mechanism for teaching data science at scale.

\section{Back Matter}

\subsection{Author Contributions}

SCH and RDP equally conceptualized, wrote and approved the manuscript. 

\subsection{Acknowledgements}

We would like to thank Elizabeth Ogburn, Kayla Frisoli, Jeff Leek, Brian Caffo, Kasper Hansen, Rafael Irizarry and Genevera Allen for the discussions and their insightful comments and suggestions on how to improve the presented ideas.


\clearpage 
\bibliographystyle{unsrtnat}
\bibliography{elements-principles.bib}

\clearpage 

\setcounter{page}{1}

\setcounter{section}{0}
\makeatletter
\renewcommand{\thefigure}{S\@arabic\c@figure}
\makeatother

\setcounter{figure}{0}
\makeatletter
\renewcommand{\thefigure}{S\@arabic\c@figure}
\makeatother

\setcounter{table}{0}
\makeatletter
\renewcommand{\thetable}{S\@arabic\c@table}
\makeatother

\onecolumn

\end{document}